\documentclass{article}
\usepackage{spconf,amsmath,graphicx}
\usepackage{multirow}

\title{IG-Track:IOU guided Siamese Networks for Visual Object Tracking}
\name{Mohana Murali Dasari,  Rama Krishna Sai Subrahmanyam Gorthi}
\address{Indian Institute of Technology, Tirupati}
\begin{document}
%
\maketitle
\begin{abstract}
Recently Deep Learning based Siamese Networks with region proposals for visual object tracking becoming more popular. These networks, while testing, perform extra computations on output if trained network, to predict the bounding box. This however hampering the precision of bounding box. In this work, the authors have proposed a network guided by Intersection Over Union(IOU) while training, to predict precise bounding box. This is achieved by adding new loss function in training the network, to maximize IOU of the predicted bounding box with ground truth. While testing on VOT2018, GOT-10k tracking benchmarks,the proposed approach outperformed the base approach by more than 10\% in terms of precision.      
\end{abstract}
\begin{keywords}
Visual Object Tracking, Deep Learning for Tracking, Siamese trackers with IOU, Region Proposals 
\end{keywords}
\section{Introduction}
\label{sec:intro}
In a video sequence, given the location of an object of interest in the first frame, finding its location in the subsequent frames can be called as "Visual Object Tracking". It has major applications including, visual surveillance, missile-guidance, weather forecasting and autonomous driving. With the advancement of deep neural networks, this computer vision task progressing very well, producing state-of the art methods.
However it is still considered as a challenge to track object under some scenarios such as, occlusion, back ground clutter, deformation, to name a few\cite{otb}. There are some benchmarks to evaluate the performance of visual object tracking, including VOT2018 \cite{vot} and GOT-10k \cite{got}.

To find the object location in subsequent frame, trackers search around the object location in the previous frame and calculate correlation with template. In Siamese network based trackers, instead of pixels, the correlation is performed between the feature maps, obtained from learned network. Some frameworks update template frequently \cite{cfnet}, while some other trackers keep it as it is in the first frame \cite{siamfc} \cite{rpn}.

Many state of the art tracking algorithms are inspired from detection framework. The SiamRPN tracker \cite{rpn}, where Region Proposal Network was introduced into tracking, was actually inspired from Faster-RCNN \cite{frcnn}, which is used for detection. In SiamRPN++ tracker \cite{rpnpp}, multiple siamrpn blocks have resemblance with cascade-RCNN\cite{casc}, which is an improved detection framework. ATOM tracker \cite{atom} has its roots in IOUNet \cite{iounet}, which is more accurate object detection framework. From these examples, we can conclude that better detection algorithms are embedded into tracking framework for enhancing the tracker performance.

Although above detection inspired trackers have obtained out-standing tracking accuracy and precision, there exist some inconsistent and redundant operations, which are hampering in utilising the full potential of these networks. Consider the ATOM tracker\cite{atom}, which is trained to predict the overlap between the target object and an estimated bounding box. For each image pair, they generate 16 candidate bounding boxes by adding Gaussian noise to the ground truth coordinates, while ensuring a minimum Intersection Over Union(IOU) of 0.1. Generating proposals by adding noise means, guessing the bounding box randomly, which is in our opinion, inconsistent with the tracking methodology. On the other hand, SiamRPN \cite{rpn} and SiamRPN++ \cite{rpnpp}trackers haven't incorporate any module to maximise IOU score during the training. While testing, these two networks perform extra operations on the output of trained network to get actual bounding box. In other words, they are not end to end trained trackers in a true sense to give bounding box in image domain. Keeping these drawbacks in the view, the authors are proposing to include IOU module on the top of Siamese Region Proposal framework training so that the new architecture can predict the bounding box,guided by maximising IOU, in the image domain.

To summarize, the main contributions of this work are listed below:
\begin{itemize}
    \itemsep0em 
    \item Guiding the training of network for tracking, by incorporating new IOU module in training 
    the network. 
    \item Adding new loss function to maximize predicted IOU, there by improving the precision.    
\end{itemize}

In this work, we have considered "SiamRPN++" as the base work.The proposed tracker, referred 
to as "IG-Track", has been trained on ImageNet VID, ImageNet DET \cite{vid_det} and COCO 
\cite{coco}. When tested on VOT2018, our tracker performance improved significantly.   

The rest of the paper is organized as follows. A brief literature review on existing frameworks are explained in Section II. The details of proposed work are presented in Section III. Section IV contains experimental details with results and analysis. Finally, this paper is concluded in Section V.

\section{Related Work}
\label{sec:format}
In this section, we introduce recent trackers, with spotlight on the Siamese network based trackers. 

The basic operation in visual object tracking involves matching the object of interest in subsequent frames, which can be converted into a point-wise multiplication operation in the frequency domain. Inspired by this, there exist many works in the literature, which considered various hand crafted features and improved tracking performance\cite{kcf},\cite{eco} while operating at above real time speed(>30FPS). With advancement of deep learning, many of these models have started incorporating features extracted from deep networks, such as AlexNet, VGG and ResNet, there by raising the performance significantly.

Recently, the Siamese network based trackers gaining popularity for their efficiency combined with accuracy. These trackers accept an image template of the object and a search image, cropped from another frame. They are fed into a network to extract feature maps, which are correlated and produce a similarity map. Although these two branches can remain fixed during the tracking phase \cite{siamfc}, some trackers such as CFNet \cite{cfnet}, updated online to adapt the appearance changes of the target. Advanced trackers like SiamRPN, DaSiamRPN and SiamRPN++ add extra modules like region proposal network, while ATOM incorporate IOU modulation and IOU prediction for improving tracking performance. 

\section{The Proposed Framework}
\label{sec:pagestyle}
\subsection{Analysis of Region Proposal based Siamese Trackers}
Inspired by the success of Siamese trackers with region proposals, we consider SiamRPN and SiamRPN++ as the base work. We are interested in developing end to end training of this tracker. In the SiamRPN, network is trained to regress bounding box in a transformed domain \cite{rpn}, which is also continued in the base work. While testing, they apply inverse transform on regression feature map, select top K proposals, impose scale penalty and aspect ratio penalty to re-rank them. These post processing steps were actually proposed in SiamFC \cite{siamfc}. On the top Non-maximum-suppression(NMS) is performed to predict the correct bounding box in the image domain. After the final bounding box is selected, target size is updated by linear interpolation to keep the shape changing smoothly.

In SiamFC tracker, where location of maximum response is alone available, using scale penalty, aspect ratio penalty for estimating bounding box width and height is applicable. However for SiamRPN and SiamRPN++ trackers, which have inherent capability to estimate bounding box width and height, these steps are not required. 

\subsubsection{Argument}
\textit{Applying the scale and ratio penalty during training of SiamRPN++ leads to over fitting.}

In preliminary experimentation, it is found that inclusion of above post processing steps are leading to over fitting of the network. This can be observed on VOT2018 dataset. For region proposal based trackers, which have capability to predict bounding box width and height inherently, imposing scale and ratio penalty are no more relevant. 
\begin{figure}[htb]
    \centerline{\includegraphics[width=8.5cm]{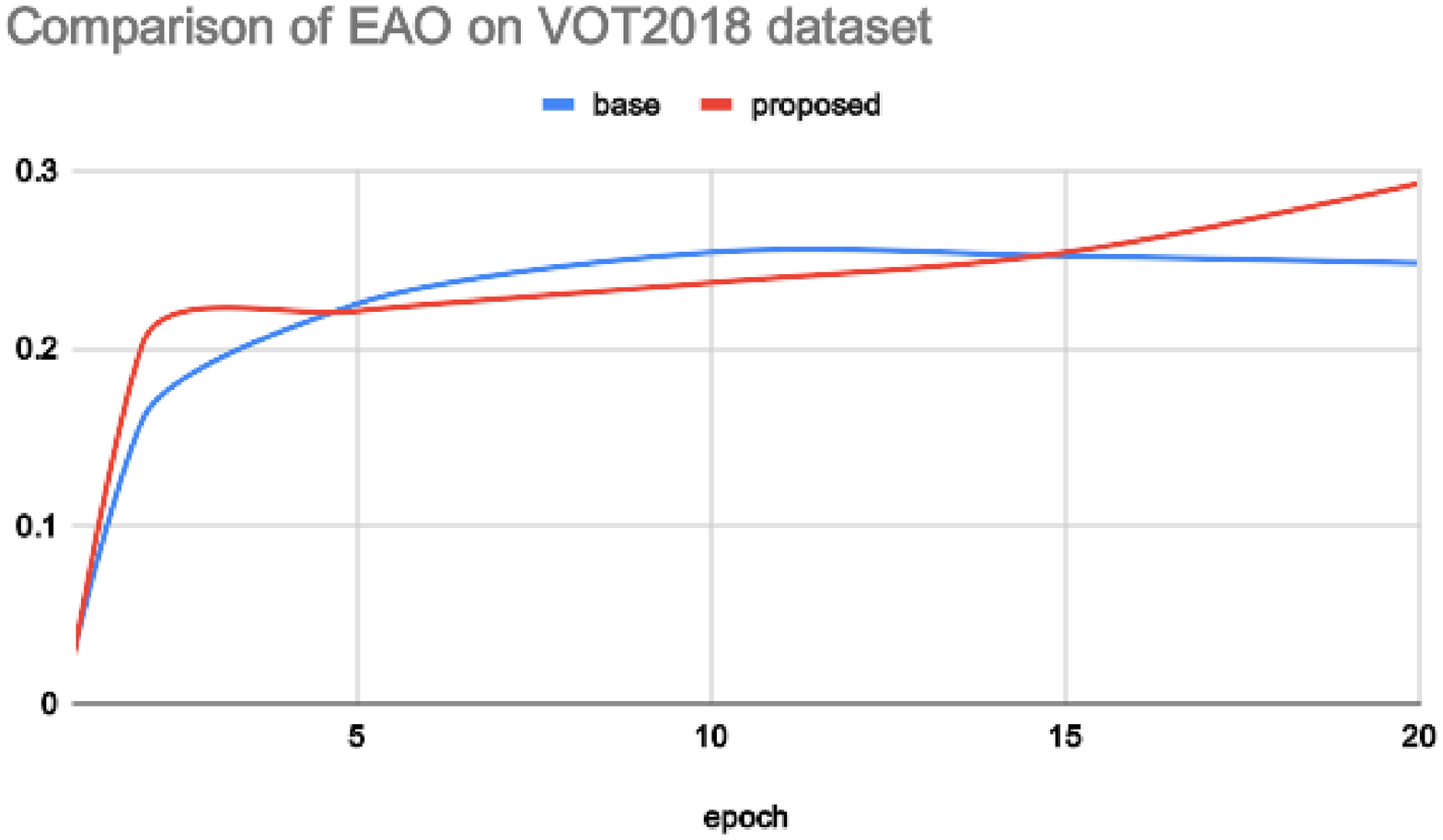}}
    \caption{Comparison of EAO with and with out ``scale and ratio penalty"}
    \label{fig:1}
\end{figure}

Hence, in this work, the authors have not included those redundant operations during the training of the proposed network for tracking.

Any good tracking framework is expected to predict bounding box that has high overlap with it's ground truth. From the deeper analysis \cite{iounet}, it is also proved that, selecting the top proposal based on classification score mostly hamper precision. Hence, to select the precise bounding box, it must be guided by Intersection Over Union(IOU) with the ground truth bounding box.

\subsection{IOU guided Siamese Networks for Visual Object Tracking}
\begin{figure*}[ht]
    \centerline{\includegraphics[width=15cm]{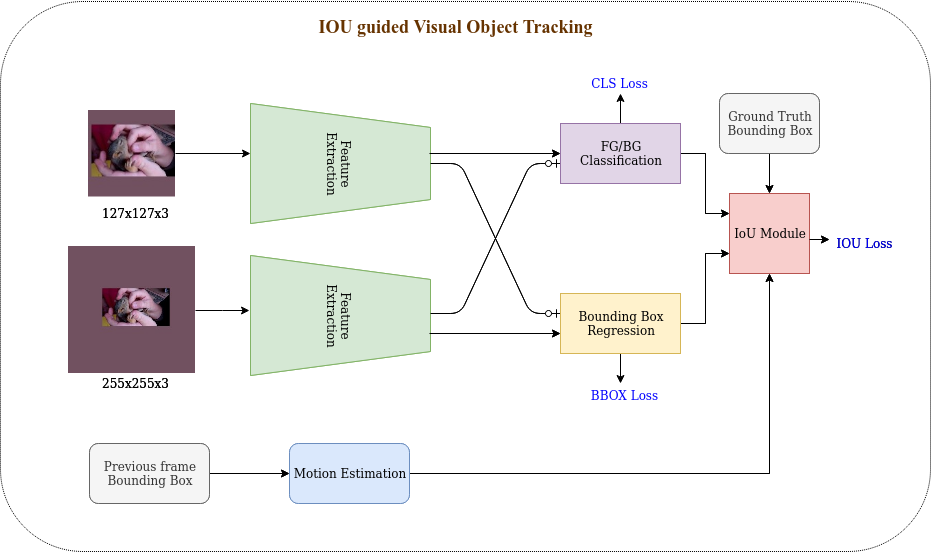}}
    \caption{Proposed IOU guided tracking frame work}
    \label{fig:3}
\end{figure*}

Using the above analysis, The authors have proposed a modified framework for end to end training of region proposal based tracker as shown in \ref{fig:3}, referred to as ``IG-Track". From the initial frame, a fixed patch(z) of 127x127 centered around the object called "Target Image", is cropped. From the next frame onward, a fixed patch(x) of 255x255 centered around the previous frame bounding box called ``Search Image", is cropped. Both these patches are passed through feature extraction network (such as AlexNet, VGGNet and ResNet50) separately for feature maps. These feature maps are then fed to classification block and regression block for calculating the corresponding losses by producing response maps in feature domain, which were originally proposed in the base tracker.

In the proposed ``IG-Track" framework, a new block called ``IOU module" is introduced during the training. It is a significant contribution of this paper. This module accepts the above feature domain response maps and convert them into image domain. From these, top-K ``probable bounding boxes" are selected with the help of anchor boxes, as done in the Inference stage in \cite{rpn}. On the other hand, IOU module also accepts a ``estimated bounding box", which is obtained from previous bounding box passed through a motion estimation block, whose parameters are also learnt by the model. IOU module then calculates the IOU between ``estimated bounding box" and ``probable bounding boxes" to produce IOU response map, whose maximum corresponds to ``predicted bounding box". For smoothness, we have used linear interpolation between previous bounding box size and predicted bounding box size. IOU loss is then defined using IOU between this predicted bounding box and ground truth bounding box. 

\begin{equation}
\label{eq1}
IOU\_loss = 1- IOU(pred\_bb, gt\_bb)
\end{equation}

As the training progress, predicted box is more aligned with ground truth, as it is guided to minimise the IOU\_loss. 

\subsection{Tracking Approach}
Tracking is performed exactly as in the base framework.

\section{Experimental Results}
\label{sec:typestyle}
\subsection{ Training and Evaluation}
\subsubsection{Training}
We have used ResNet50 architecture for feature extraction. Before training, the feature extraction network is loaded with pre-trained weights. Network is then trained using Imagenet VID, Imagenet DET and COCO.  
\subsubsection{Evaluation}
The proposed framework is evaluated on benchmark data-sets including VOT2018, OTB2015 and GOT-10k. 
\subsection{Implementation Details}
The training loss is combination of three losses i.e. classification loss, bounding box regression loss as defined in \cite{rpn} and additionally introduced IOU loss in Eq (\ref{eq1}). 
\begin{equation}
    Loss = L_{cls} + L_{reg} + L_{iou} 
\end{equation}
It is trained for 40 epochs with exponentially decaying learning rate from 0.005 to 0.00005. System has 4 GPU cards, each having a NVIDIA GeForce GTX Ti-1080 with 4GB RAM. All parameters kept as in the base framework. 

\subsection{Comparison with the Base Tracker}
It is well known that given more data for training, deep learning based models likely to improve  the performance. In the base work, the authors of SiamRPN++ \cite{rpnpp} have trained their network with four data-sets including YouTube-BB. This data-set consist of 0.24 Million videos and its impact is so high that, in case of SiamRPN, training additionally with Youtube-BB boosts VOT2016’s EAO from 0.317 to 0.344 \cite{rpn}. Hence, for fair comparison, we have re-trained the base tracker using the above three data-sets only(VID, DET and COCO) and used this alone for comparison with the proposed framework. As ATOM tracker \cite{atom} is trained on entirely different data-sets such as LaSOT and TrackingNet, which are too huge, it is not used for comparison. The objective of this study is only to improve the training base tracker.  

\subsubsection{VOT2018 Dataset}
The VOT-2018 dataset, one of the latest benchmark available for evaluating short term single object trackers, consist of 60 videos with different challenges. To evaluate different trackers, they use metrics such as Expected Average Overlap, Accuracy and Robustness. The comparison of base tracker SiamRPN++ with proposed IG-Track tracker on VOT-2018 dataset is reported in the Table \ref{tbl_vot}. 

\begin{table}[ht]
\caption{Performance comparison on VOT-2018}
\begin{center}
\begin{tabular}{|c | c | c | c |} 
 \hline
   & Base & Proposed & Improvement \\ [0.5ex] 
 \hline\hline
 EAO $\uparrow$ & 0.290 & \bf{0.327} & 13\%  \\ 
 \hline
 Accuracy $\uparrow$ & \bf{0.571} & 0.565  & -1\%\\
 \hline
 Robustness $\downarrow$ & 0.347  & \bf{0.309} & 11\%\\ 
 \hline
\end{tabular}
\end{center}
 \label{tbl_vot} 
\end{table}

From the table it can be observed that, the proposed tracker have significant improved performance in terms of EAO(12\%) and Robustness(11\%) while maintaining almost same Accuracy. It can be attributed to the introduction of IOU module with IOU loss in the proposed method, which can improve the precision of tracking.

\subsubsection{GOT-10k Dataset}
The GOT-10k dataset, a large high-diversity benchmark for Generic Object Tracking in the Wild, consist of 180 videos with 84 object classes. To evaluate different trackers, they use metrics such as Average Overlap(AO), Success Rate SR($_{0.5}$) and SR($_{0.75}$). The comparison of base tracker with proposed tracker on GOT-10k dataset is reported in the Table \ref{tbl_got}. 

\begin{table}[ht]
\caption{Performance comparison on GOT-10k}
\begin{center}
 \begin{tabular}{|c | c | c | c |} 
 \hline
   & Base & Proposed & Improvement \\ [0.5ex] 
 \hline\hline
 AO $\uparrow$ & 0.453 & \bf{0.459} & 1\%  \\ 
 \hline
 SR$_{0.5}$ $\uparrow$ & 0.546 & \bf{0.558}  & 2\%\\
 \hline
 SR$_{0.75}$ $\uparrow$ & 0.195  & \bf{0.220} & 12\%\\ 
 \hline
\end{tabular}
\end{center}
 \label{tbl_got} 
\end{table}

From the table it can be observed that, the proposed tracker have significant improved performance in terms of SR$_{0.75}$(12\%), which measures the percentage of successfully tracked frames where overlap rates are above 0.75. This clearly shows the importance of IOU module and IOU loss in the proposed framework. 






\section{Conclusion and Future Work}
In this paper, the authors have proposed an end to end trainable tracking framework, named as IG-Track, which can be used for visual object tracking. It is showed that, precision of state of the art SiamRPN++ tracker can be improved by incorporating extra block called IOU module and back propagating IOU loss. Especially on the VOT2018 and GOT-10k datasets, IG-Track out performed the base tracker in terms of precision. In future, there is a scope for improving the accuracy and precision by modelling the motion using Recurrent Neural Networks.

\bibliographystyle{IEEEbib}
\bibliography{strings,refs}

\end{document}